# Title: Earthquake and Electrochemistry: Unraveling the Unpredictable

**Author:** Atanu Das[1, *]

**Affiliations:**

[1]Institute of Applied Mechanics, School of Aeronautics and Astronautics, Zhejiang University, Hangzhou-310027, Zhejiang Province, Peoples' Republic of China

*Correspondence to: atanu@gmx.com (Atanu Das*); atanu.esensex@protonmail.com

## ABSTRACT

Earthquakes are measured using well defined seismic parameters such as seismic moment ($M_o$), moment magnitude ($M_w$), and released elastic energy(E). How this tremendous amount of energy is accumulated silently deep inside the earth's crust? The most obvious question in seismic research remains unanswered. We found an inherent and intriguing connection between the released energy in an earthquake and electrochemical potential induced in an ultra-thin metal oxide electrode immersed in an aqueous pH solution, which leads us to understand the origin of the energy accumulation process in an earthquake.

A huge electrochemical potential is accumulated from numerous electrochemical cells formed in a unique layer structure of hydrated clay minerals (predominantly smectite), which resulted in a lightning-like discharge in the lithosphere (hypocenter). The subsequent thunder-like massive shockwave is produced, which initiates tectonic plate movement along a fault line, probably through acoustic fluidization (AF), and resulting seismic energy is transmitted as primary wave (P-wave), secondary wave (S-wave), and surface waves.

The presence of electrical voltage in the hypocenter directly supports the seismic electric signal (SES), further strengthening the VAN method of earthquake prediction. Our finding is supported by a plethora of research and observation devoted to seismic science. This study will indeed find its significance if immediate action is implemented to monitor the evolution of electrochemical potential, seismic electrical signal (SES), and ionic activity in the fault zone at lithosphere as well as in the ionosphere for predicting an impending earthquake for saving human lives as early as possible.

## 1. INTRODUCTION

Earthquake is the most complex natural calamity claiming human lives and mass destruction of manmade structures. A natural earthquake occurs due to a sudden slip or collision of tectonic plates in a fault zone deep in the earth's crust, and the stored energy is radiated as seismic waves causing ground motion and acceleration. With tremendous progress of modern science and instrumentation, seismometer and earthquake source theories are matured enough to provide quantitative source parameters than the only magnitude. A well-established moment magnitude ($M_w$) scale (*1, 2*) has been used to quantify shallow and deep earthquakes based on radiated wave energy. This magnitude scale derives from the concept of the seismic moment ($M_0= \mu AD$), which is directly related to the measurable parameters in a seismic event such as the area of the rupture(A) along the geological fault, the average of displacement (D) or slip during the rupture and shear modulus (μ) or elastic constant. The famous Richter-Gutenberg energy-moment magnitude

relation(*3*) of a seismic event is Log E=4.8+1.5M$_w$ where E is associated with elastic energy in Joule which drives the earthquake to progress from the hypocenter (foci), M$_w$ is the moment magnitude of an earthquake. How this tremendous amount of energy is accumulated deep in the earth's crust? We have no clues on the origin of such an energy accumulation process occurring silently inside deep in the earth's lithospheric region. Most of the seismic study only quantify parameters after an actual seismic event occurred. The ultimate goal of earthquake research is to understand the seismic source's nature and a timely forecast of an impending earthquake by finding suitable pre-seismic precursors.

Significant geochemical precursors, i.e., the anomalous concentration of dissolved ions(*4*) and gases in groundwater, have been measured before intermediate and large earthquakes. Radon ($^{222}$Rn) gas anomaly in the environment (soil) and groundwater before the earthquake was detected in many cases(*5*). A wide range of pre-earthquake phenomena has been reported involving ground and satellite-based observations. Atmosphere-ionosphere response a few days before the M9 great Tohoku Japan earthquake revealed a rapid increase of emitted outgoing longwave radiation (OLR) of infrared (10-13 µm) and increased in a variation of total electron content (TEC) in the ionospheric region above the epicenter(*6*). Thermal infrared (TIR) and ionosphere anomaly was detected in the Iran earthquake (*7*), 2001 7.6 Bhuj earthquake(*8*), and 2017 M$_w$6.5 Jiuzhigou earthquake (*9*). Such pre-seismic anomalies are possibly linked to the energy accumulation process before the actual earthquake strike. The most debated and criticized VAN method(*10*) has been used in Greece and other countries for earthquake prediction. The method is based on detecting geo-electric potential or seismic electrical signals (SES) that appear prior to earthquakes.

While working on an electrochemical sensor, we found a gradual increase of electrochemical potential in a solid-state electrode, which reminds us of the gradual increase of energy with an earthquake's magnitude. What a connection and consequences! We found a close match between estimated electrochemical potential induced in a solid-state electrode and energy released in an earthquake while looking more deeply into earthquake energy release patterns corresponding to the magnitude scale. With this clue, we investigated the origin of the energy accumulation process through an electrochemical reaction involving clay minerals and water.

## 2. RESULTS
**2.1 Derivation of Electrical Potential from Released Energy in an Earthquake**

Richter-Gutenberg energy-magnitude provides an accurate estimation of seismic energy corresponding to the moment magnitude scale.

$$Log E = 4.8 + 1.5 M_w \tag{1}$$
$$E = 10^{(4.8+1.5M_w)} \text{ (Joule)} \tag{2}$$

E is the released energy in Joule, and M$_w$ is the moment magnitude of an earthquake event. The magnitude M$_w$= 0 in the moment-magnitude scale has the equivalent energy of 63.09573×10$^3$ Joule. The electrical potential could be derived from released energy from the simple calculation. The electrical potential can be calculated as follows,

$$63.09573 \times 10^3 \text{ J} = 0.01752 \text{ kWh } [1 \text{ J} = 2.77778 \times 10^{-7} \text{ kWh}] \tag{3}$$
$$= 0.01752 \times 3600 \times 10^3 \text{ V A s}$$
$$= 0.01752 \times 3600 \times 10^3 \text{ V C}$$
$$63.09573 \times 10^3 \frac{\text{J}}{\text{C}} = 0.01752 \times 3600 \times 10^3 \text{ V} \tag{4}$$
$$= 0.01752 \times 3.6 \times 10^6 \text{ V}$$
$$= \text{SEP}_{M_w=0.0} \times 3.6 \times 10^6 \text{ V}$$

We assign SEP as seismic electrical potential, which only increases exponentially with the increase of magnitude scale. Another factor, i.e., $3.6 \times 10^6$ will remain constant in $M_w$ (0~1.9). Similarly, we can calculate the electrical potential for other earthquake moment magnitudes. **Table 1** shows details of earthquake magnitude $M_w$ (0~1.9) and corresponding electrical potentials. The magnitude scale $M_w$ (0~1.9) is considered as an initial basic set for calculation. Other sets like $M_w$ (2~3.9), $M_w$ (4~5.9), $M_w$ (6~7.9), $M_w$ (8~9.9) will follow (see **Supplementary Materials**, **Table S1~S4**) the same trend of $M_w$ (0~1.9) with increasing multiplication factor. One may simply correlate such huge electrical potential because of the fault zone's energy accumulation process before an actual earthquake strike. The following section will reveal an intriguing connection between earthquake electrical potential and potential originated from a spontaneous electrochemical reaction.

## 2.2 Electrochemical Potential at the Electrode-Aqueous Interfaces

The metal-oxide electrode responds to the aqueous pH buffer solution through a reversible electrochemical reaction. The redox electrochemical reaction occurring at the electrode-aqueous interface can be written as,

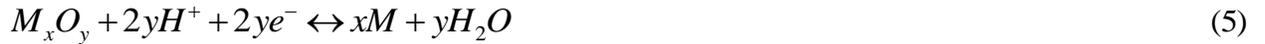
$$M_xO_y + 2yH^+ + 2ye^- \leftrightarrow xM + yH_2O \tag{5}$$

The electrode potential for the above reaction from the Nernst equation is,

$$E = E^0 - \frac{0.0591}{2y}\log[a_{H^+}]^{2y} = E^0 - (\frac{2y}{2y}) \times 0.0591 \log[a_{H^+}] = E^0 - 0.0591 \times pH \tag{6}$$

where $E^o$ is the standard cell potential, 0.0591 V is the Nernst potential at 25°C.

A. Das et al. (11) recently proposed a generalized Nernst equation with asymmetric ion exchange for ultra-thin metal oxide system. A generalized redox reaction involving asymmetric ion exchange can be written as,

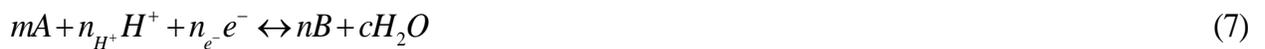
$$mA + n_{H^+}H^+ + n_{e^-}e^- \leftrightarrow nB + cH_2O \tag{7}$$

where A may represent a simple metallic ion and B the corresponding metal.
The electrode potential for the above reaction is given below:

$$E = E^0 - \frac{0.0591}{n_{e^-}} \log \frac{(a_B)^n (a_{H_2O})^c}{(a_A)^m (a_{H^+})^{n_{H^+}}} = E^0 - 0.0591 \left(\frac{n_{H^+}}{n_{e^-}}\right) \log(a_{H^+}) - \frac{0.0591}{n_{e^-}} \log \frac{(a_B)^n (a_{H_2O})^c}{(a_A)^m} \tag{8}$$

Assuming the activity of A, B, and H₂O to unity, A generalized Nernst equation for the electrode potential is,

$$E = E^0 - 0.0591 \left(\frac{n_{H^+}}{n_{e^-}}\right) \log(a_{H^+}) = E^0 - 0.0591 \times x \times pH \tag{9}$$

The $n_{H+}$ is the number of moles of hydrogen ion chemisorbed (chemical reaction) at the electrode surface, and its values are 1, 2, 3, …, n. The $n_{e-}$ is the number moles of electron transfer, which is the same as the change of oxidation number of the material ($\Delta Z$) involved in the redox (A→B) reactions. Therefore, 'x' has a precise stepwise discrete value, which controls the ultrathin oxide electrode's potential in a similar discrete manner. The ion exchange factor '$x= n_{H+}/n_{e-}$' will determine the exact electrochemical cell potential in a spontaneous electrochemical reaction with an aqueous solution. We assign NP as the Nernst potential (0.0591 V $pH^{-1}$), which is a special case of equation (9) owing to symmetric ion-exchange factor ($x=1$). Electrode potential due to asymmetric ion exchange ($x\neq 1$) can be assigned as Pourbaix Potential or PP as the idea of asymmetric ion exchange was originally formulated by M. Pourbaix through pH-potential formulation(*12*). We could relate PP and NP by a simple equation,

$$PP_x = NP \times x \qquad (10)$$

The electrode potential could be expressed as

$$E = E^0 - PP_x \times pH \qquad (11)$$

E is electrochemical potential due to half-cell reaction occurring at the working electrode-aqueous interface. Another half-cell reaction ($E^o$) is occuring at the reference (Ag|AgCl) electrode, which is well understood non-interfering rection. The half-cell reaction of each working electrode could be added as a regular electrochemical cell (See **Supplementary Materials**, **Figure S1**). For N number of electrochemical cells ($N_{Cell}$), the total electrochemical potential will be,

$$N_{Cell} \times (E-E_0) = PP_x \times pH \times N_{Cell} \qquad (12)$$

The abundance of $SiO_2$ and $Al_2O_3$ in the earth crust leads us to focus on materials with $\Delta Z=+3, +4$) and associated electrochemical potential in aqueous solutions with different ionic exchange factors. In laboratory condition, the various electrochemical potential for $Al_2O_3$ and $SiO_2$ was reported. The minimum value for $Al_2O_3$ was 17 mV for low temperature grown oxide(*13*) whereas the ideal value of 40-55 mV is expected(*13*). Reddy et al. reported 30 mV $pH^{-1}$ for their porous silicon (P-Si) structure(*14*). Naif et al. (*15*) reported 66 mV $pH^{-1}$ in a similar P-Si sensing electrode. Based on Pourbaix formulation, ideal Si-$SiO_2$ could exhibit electrode potential of 59.16, 73.9, and 88.6 mV depending on the ion-exchange factor 4/4, 5/4 and 6/4, respectively. The details of electrochemical reactions and associated electrode potentials are described below.

$\Delta Z = +4 (0 \to +4)$

$$Si + 2H_2O \leftrightarrow SiO_2 + 4H^+ + 4e^- \quad \Rightarrow E = E^0 - 0.0591 \times pH \qquad (13)$$

$$Si + 3H_2O \leftrightarrow H_2SiO_3 + 4H^+ + 4e^- \quad \Rightarrow E = E^0 - 0.0591 \times pH + 0.0148 \log(H_2SiO_3) \qquad (14)$$

$$Si + 3H_2O \leftrightarrow HSiO_3^- + 5H^+ + 4e^- \quad \Rightarrow E = E^0 - 0.0739 \times pH + 0.0148 \log(HSiO_3^-) \qquad (15)$$

$$Si + 3H_2O \leftrightarrow SiO_3^{2-} + 6H^+ + 4e^- ; \quad \Rightarrow E = E^0 - 0.886 \times pH + 0.0148 \log(SiO_3^{2-}) \qquad (16)$$

Very high electrode potential 300 mV $pH^{-1}$ was achieved in 3D nano-porous P-Si structure(*16*) which could be due to ion exchange factor of $x=21/4$ ($PP_{21/4}=310$ mV). It is indeed a beautiful example of interfacial engineering for achieving high electrode potential in Si-$SiO_2$ system. Trend of increasing electrode potential from 30 mV to 300 mV for a unit electrochemical cell in Si-$SiO_2$ system clearly indicates that the value could be much more in case of extreme engineered 3D surfaces.

## 2.3 Correlation between Electrochemical Energy and Elastic Seismic Energy

Based on the experimental evidence of electrode potential in Si-SiO$_2$ system and Pourbaix formulation of higher-order ion-exchange in several materials system (Ga-Ga$_2$O$_3$, Ir-Ir$_2$O$_3$, Cr-Cr$_2$O$_3$, V-V$_2$O$_3$, and many more), it seems logical to generalize the Pourbaix Potential for any material system. We could extrapolate the trend of higher-order ion exchange by considering chemisorption of an increasing amount of hydrogen ion (H$^+$) in the electrical double layer (EDL) formed at the electrode/aqueous interface. Considering the lowest order ion-exchange factor x=1/4 for Si-SiO$_2$ and x=1/3 for Al-Al$_2$O$_3$, we calculated the PP for both materials.

$$PP_{x=1/4} = 0.05916 \times \frac{1}{4} = 0.01479 \tag{17}$$

$$PP_{x=1/3} = 0.05916 \times \frac{1}{3} = 0.01972 \tag{18}$$

At this point, we found a striking similarity between seismic electrical potential (SEP) and Pourbaix Potential (PP). In fact, the average of PP$_{x=1/4}$ and PP$_{x=1/3}$ is 0.017255, which is very close to the SEP$_{Mw=0.0}$=0.01752. Other factors from equation (12) i.e., pH×N$_{Cell}$ could be seen as equivalent of 3.6×10$^6$ of equation (4). Pourbaix Potential for different ion-exchange factor for Al-Al$_2$O$_3$ (ΔZ= +3) and Si-SiO$_2$ (ΔZ=+4) are presented as separate MS Excel datasheet (see **Supplementary Materials**-External Database). After careful observation, we choose several values closely match with seismic electrical potential (SEP) (see **Table 1**). Selected Pourbaix potential is shown in **Table 2**. To understand the correlation between SEP and PP, both SEP (V) vs. M$_w$ (0.0~1.9) and Log (SEP) vs. M$_w$ (0.0~1.9) curves (**Figures 1A and 1B**) are plotted accommodating the closely matched PP values from **Table 2**. All curves are surprisingly overlapped, which indicates an inherent link between SEP and PP. The theoretical temperature effect on the electrode potential is 0.02 mV per degree Kelvin(*17*). Considering temperature of 200°C at 75 km depth(*18*) in the earth crust, the change of electrode potential will be ΔT×0.02 mV=175×0.02 mV=3.5 mV, which is very small and does not significantly alter the original Pourbaix Potential calculated at 298.15 K(25°C). The details of PP values at 200°C are presented in **Table S5** (see **Supplementary Materials**) and the corresponding plot in **Figures S 2A and 2B**. Evolution of PP from PP$_{1/4}$ =0.01725 V (P$_{1/3}$=0.01972 V) to PP$_{839/4}$ =12.408 V (PP$_{629/3}$ =12.403V) in an exponential manner could be explained through the Elovich model of chemisorption kinetics. The Elovich equation has been widely used in adsorption kinetics, which describes the chemisorption (chemical reaction) mechanism in nature through a multilayer adsorption site(*19-21*). According to this equation, the adsorption site increases exponentially with adsorption. It is noteworthy to mention that the maximum Pourbaix potential for each cell is 12.40 V, and the value is linked with two fundamental constants, i.e., Planck constant and speed of light in a vacuum (hc=12.39 eV×100 nm). We believe that this the fundamental limit of hydrogen ion adsorption in the electric double layer (EDL) of electrode/aqueous interface. A compact table (**Table 3**) is prepared to show the moment magnitude scale and corresponding SEP or PP and pH×N$_{Cell}$ values. As the pH of water usually varies from 1~12 in pH scale, the number of electrochemical cells (N$_{Cell}$) will vary accordingly, keeping the total electrical potential constant as per the moment magnitude-energy relation. At this point, we strongly believe that the electrochemical potential accumulated from numerous electrochemical cells is the origin of earthquake energy. This electrical energy transforms into mechanical energy (seismic energy) to drive the entire earthquake process.

# 3. DISCUSSION

Originally earthquake energy is calculated from the concept of stored elastic energy in the rock-forming materials. The extracted seismic electrical potential (SEP) closely match with Pourbaix Potential (PP). Is this an accidental coincidence, or are they inherently linked? How to validate an utterly different origin 'a huge electrochemical potential' as an earthquake triggering source? We strongly believe that the accumulation of electrical potential and electric field in the earth's crust prior to any earthquake event could provide a unified model to explain most of the existing observation and findings devoted to seismic sciences unequivocally. "Extraordinary Claims Require Extraordinary Evidence" (ECREE), a famous phrase was made popular by astronomer Carl Sagan. This aphorism is at the heart of scientific skepticism, a model for critical and rational thinking. In the following sections, we will try to validate our claim by systematic analysis and logical clarifications of related findings directly or indirectly linked to our study.

## 3.1 Origin of Electrochemical Potential in Fault Zone and Role of Clay Minerals

The electrochemical potential obtained from redox reaction of $Al_2O_3$ ($\Delta Z=+3$) and $SiO_2$ ($\Delta Z=+4$) with aqueous pH buffer in different ion exchange factor closely match with the earthquake electrical potential (SEP). Naturally, we must focus primarily on the source of the $SiO_2$ and $Al_2O_3$ in the earth's crust and earthquake fault zone. In nature, during the weathering process, rocks react with water and produce clay minerals. Clay minerals are phyllosilicates, a stacked layer of two dimensional (2D) sheets of hydrated aluminosilicates ($SiO_2$, $Al_2O_3$, $H_2O$) found in geologic deposits, terrestrials weathering environment, and marine deposits(*22*). Water is present in a variable amount as part of the structure of minerals. Based on the stacking of silicon tetrahedra (T) and alumina octahedra (O) sheets in the crystal unit or layers, clays are categorized into three different groups, i.e., 1T:1O, 2T:1O, 2T:2O type clay minerals. A schematic representation of different clay minerals is shown in **Figure 2**. Clay minerals have a wide range of cation-exchange capacity (CEC) due to the crystal lattice's overall charge imbalance. This property arises when either Si (IV) or Al (III) is substituted with a lower valent metal cation such as Fe (III) or (Al (III) or Mg (II)), respectively, resulting in a net negative charge region in silicate layer.

Kaolinite ($2SiO_2Al_2O_3 2H_2O$) is the most prominent 1T:1O clay mineral (**Figure 2A**). It is theoretically composed of 47% $SiO_2$, 39% $Al_2O_3$, and 14% $H_2O$. The layers are held together with strong hydrogen bonding. Interstitial cations and water do not enter between the structural layers mineral particle when the clay is wetted. The effective surface of kaolinite group minerals is restricted to its external surface for an interfacial reaction. Smectite minerals are 2T:1O type, which is characterized by the octahedral sheet (1O) sandwiched between two tetrahedral (2T) sheets (**Figure 2B**). It is theoretically composed of 67% $SiO_2$, 28% $Al_2O_3$, and 5% $H_2O$ (without interlayer water). Owing to the weak van der Waals force between interlayer, waters, and exchangeable cations ($K+$, $Ca+$, and $Mg^{2+}$) are easily entered into the interlayer spacing of smectite, causing expansion and swelling when the clay is wetted. Vermiculites are also 2T:1O type minerals with very high negative charge associated with these minerals owing to tetrahedral substitution by a lower valent metal ion (**Figure 2C**). Water molecules with other cations are strongly adsorbed in the interlayer spacing. The cation exchange capacity (CEC) of vermiculite is higher than all other silicate clays, including smectite, because of the very high negative charge in the tetrahedral sheet. The internal surface exceeds the external surface of clay crystal for both smectite and vermiculite.

Illite is a non-expanding (2T:1O) clay minerals, where potassium ion ($K+$) acts as a binding site preventing the expansion of crystals (**Figure 2D**). It is theoretically composed of 12% $K_2O$, 45% $SiO_2$, 38% $Al_2O_3$, and 5% $H_2O$. The hydration, cation adsorption is less intense in fine-grained illite than in smectite but more than kaolinite due to the presence of interstratified layers of smectite or vermiculite. Chlorites are 2T:2O type non-expanding clay, which is iron magnesium silicate with some aluminum present. The crystal unit contains two silica tetrahedra sheet (2T) and two magnesium dominated octahedral (2O) sheet (**Figure 2E**). The effective surface of illite and chlorites are also restricted to its external surface for interfacial reactions. A comparison of essential properties is presented in **Table 4**(*23*).

In reality, specific clay mineral does not occur independently; instead, it is common to find several clay types in an intimate mixture. Such minerals are known as a mixed layer or interstratified such as "chlorite-vermiculite, "mica - smectite," etc. The clay mineral structure is fascinating with attractive and large surface areas for interfacial reactions in aqueous environments. In general, it is estimated that a cubic centimeter (1$cm^3$) of clay has a reactive surface of around 2800 $m^2$, which is equivalent to the area of a football field! For an analogy, a centimeter-thick pad of paper includes about 100 sheets, whereas a centimeter-thick layer of clay minerals includes about 100,000 sheets. Such a huge number of clay sheets gives us a sense of the number of $N_{Cell}$ required to satisfy the values in **Table 3**. Smectite and vermiculite will be the perfect candidate for harnessing such huge electrochemical potential owing to cation exchange capacity in the interlayer spacing and outstanding adsorption properties. Dissociative chemisorption of water(*24*) on silica surface and formation of hydronium ion could explain the increasing number of hydrogen ion adsorption in clay surface. It seems logical to think that each interlayer tetrahedral surfaces may serve as one active electrochemical cell separated by an octahedral sheet. The concept of generation of electrochemical potential in unique layer (surface) structure of clay minerals is based on several previously reported experimental results of Al-$Al_2O_3$/aqueous and Si-$SiO_2$/aqueous system together with Pourbaix formulations of pH-Potential diagram of Si-$SiO_2$ and Al-$Al_2O_3$ system.

The type of clay minerals found in the scientific deep drilling project supports our claim. The Taiwan Chelungpu-fault Drilling Project (TCDP) revealed a spike in smectite(*25*), decreases in illite, and disappearance of chlorite and kaolinite in the primary slip zone (PSZ). Integrated Ocean Drilling Program Expedition 343 (Japan Trench Fast Drilling Project, JFAST) was carried out after one year of the 2011 Tohoku-Oki earthquake ($M_w$ 9.0). The mineralogical analysis revealed that the shallow portion of megathrust is enriched in smectite (60-80 wt.% ) compared to the surrounding sediments(*26*). Scientific drilling of the San Andreas Fault also revealed weak mineral smectite(*27*).

### 3.2 Electrochemistry in Clay mineral-aqueous interface and Role of Water in the Earthquake Triggering

We realized that the electrochemical reaction between clay minerals and water is the source of electrical potential, which slowly buildup and finally trigger an earthquake. Our claim is supported by several observations and scientific studies where water plays a vital role in triggering an earthquake. Reservoir induced seismicity (RIS) is an active area of seismic research to figure out the role of water and water pressure in earthquake triggering. Gupta et al. (*28*) have reported an extensive review of the artificial water reservoir's role in triggering earthquakes. Damaging earthquake exceeding M6 occurred due to the Hsinfengkiang reservoir, China, in 1962. The devastating earthquake M7.9 in Sichuan in May 2008 may have been linked to the huge Zipingpu Dam as the epicenter is very close to the dam. On December 10, 1967, the Koyna (India) earthquake of M6.3 was another damaging reservoir triggered earthquake. We strongly believe that the reservoir water slowly diffuses to the underneath clay minerals of the nearby fault zone and create a perfect condition for a gradual buildup of electrochemical potentials for triggering an earthquake.

Water injection to stimulate hot and dry rock of the proposed geothermal project in Basel, Switzerland, triggered excessive seismic activity putting the project on hold(*29*). The possibility of the tidal triggering of large earthquakes has been extensively studied to find the correlation between the earthquake and tidal stress(*30*). A tidal variation on earth's water occurs in each new moon and full moon days due to the combined gravitation pull of moon and sun. The enhanced gravitational pull on the water body near the fault zone will increase the possibility of electrochemical reactions which could be directly linked to earthquakes. Spike in earthquake following the typhoon Morakot(*31*), Taiwan could also be linked to our findings. Himalayan earthquake is less frequent in the monsoon season, but the number increases significantly in the winter. Michas et al. (*32*) recently reported the correlation between seismicity and water level fluctuation in Polyphyto Dam, North Greece. We believe that water level variation could create suitable conditions for electrochemical reactions.

### 3.3 Evidence of Electrical Potential
Accumulation of electrical potential is directly linked to the pre-seismic precursor observed and reported in many literatures.

### 3.3.1 Ionospheric Perturbations
Planet earth works as a fabulously complex system, and the lithosphere is inherently connected to the atmosphere. We may not know how the earthquake affects the atmosphere. Variation in ionospheric Total Electron Content (TEC) was observed before many earthquakes ($>M_w 5.0$) through the analysis of global positioning satellite data. Lithosphere-Atmosphere-Ionosphere coupling (LAIC) is a well-known model for understanding the perturbation in the ionosphere. The appearance of geo-electric voltage in the seismically active region is believed to be one of the possible sources of ionospheric perturbations(*33*). Pulinets et al. (*34*) proposed that due to anisotropy of atmospheric conductivity at a height greater than 60 km, the large-scale high-intensity vertical electric field appearing at a seismically active region few days before mainshock can penetrate the ionosphere and create ionospheric anomalies. TEC anomaly is usually observed above large thunderstorms, whereas no detectable localized TEC variation is observed for a thunderstorm-quiet night(*35*).

### 3.3.2 Terrestrial Outgoing IR Emission
Emission of infra-red radiation of wavelength 8-13 µm was detected(*36, 37*) from ground-based observation satellite before several large earthquakes. The electrical potential and associated electrical field before any earthquake will excite the $SiO_4$ tetrahedral stretching electronic bond of clay minerals or rock-forming silica, which could emit huge infrared radiation typically in the range 800–1300 cm$^{-1}$ (7.7–12.5µm) (*38, 39*).

### 3.3.3 Ground Water Chemistry, Radon Gas Emission and Earthquake Light
Evidence of variation of dissolved species in groundwater before an earthquake was observed and reported in many literature(*4*). Radon gas anomalies in groundwater and soil gas was studied aiming to predict earthquake(*5*). Earthquake light (EQL) was observed just before or during strong earthquake events(*40*). Electrical field-induced ionization of different crustal elements could easily dissolve in groundwaters, leading to a change of groundwater chemistry. A trace amount of Uranium or Uranium oxide is always present in the rock-forming materials(*41*). Electronic configuration of Uranium could be written as [Rn] 5f3 6d1 7s2 which has six valence electrons. A strong electrical field could accelerate the spontaneous decay process of Uranium and facilitate the formation of more stable Radon, which we usually detect before an impending earthquake. Similarly, the electric field-induced ionization of environmental gases like nitrogen, oxygen could explain the strange luminous phenomena 'EQL' observed and recorded during strong earthquakes.

### 3.4 High Energy Electrical Discharge and Generation of Acoustic Shock Wave

The electrical potential will continue to rise exponentially in an earthquake preparation zone until it reaches its maximum electrical breakdown voltage of the surrounding materials/medium. Naturally, a high energy electrical discharge will occur in the medium, which produces an acoustic shockwave. For an analogy, natural lightning and thunder will better visualize this initial phase of the earthquake. Thunder like roaring (rumble) sound was heard immediately prior to the onset of felt vibrations caused by an earthquake(*42*). Sometimes the sound is compared with an explosion or sound of a freight train (see **Supplementary Materials, additional information**). Sometimes no shock was reported after an intense roar was noticed(*43*). Sample collected from PSZ [Taiwan Chelungpu-fault Drilling Project (TCDP)] shows features relating to melting or amorphous materials under SEM and TEM observations(*25*), which could be a signature of lightning-like discharge. Similar amorphous rock fulgurites(*44, 45*) are formed due to lightning strikes on the quartz-rich soil/mountain. Okazaki et al. (*46*) studied lawsonite dehydration as an earthquake triggering source in subducting oceanic crust. This dehydration could be resulted due to the immense heat produced during electrical discharge. It is generally accepted that weak fault materials cannot accumulate a large amount of elastic strain energy. The presence of smectite or interstratified weak clay minerals in the primary seismic slip zone excludes the possibility of storing huge elastic strain energy for earthquake triggering. Acoustic fluidization (AF) is the most plausible physical process for dynamical fault weakening(*47-49*). Fault starts to slip as long as the high-frequency acoustic shock waves remain sufficient to lead AF before dissipating within low-acoustic impedance fault gouge(*50, 51*). Thus, AF induced mechanical fault-slip radiates seismic waves (P-wave, S-wave, and surface wave). Lin et al. (*52*)studied naturally occurring thunder as a seismic source. The acoustic shockwave from strong thunder could be coupled to the ground and excite P and S waves near the surface, causing seismicity(*53*).

Finally, a schematic illustration is presented in **Figure 3** to describe the whole earthquake process based on the previous and present understanding.

## 4. CONCLUSIONS

We have derived electrical potentials from seismic energy in the moment magnitude $M_w$ (0.0-1.9) scale. We found a close match between seismic electrical potential (SEP) and Pourbaix Potential (PP) which leads us to find an electrochemical origin of earthquake energy. Weak clay minerals found in several scientific deep drilling projects excludes the possibility of stored strain energy as an earthquake triggering source. We have identified smectite and vermiculite as strong candidates for the generation of substantial electrochemical potential due to its excellent adsorption and cation exchange capacity in the silicon tetrahedral-aqueous interface. The exponential increase of Pourbaix Potential is supported by the Elovich model of chemisorption kinetics in clay minerals' multilayer adsorption site. We strongly believe that huge electrochemical potential is the possible earthquake triggering source that drives the entire earthquake process through lightning-like discharge and thunder like shockwave propagation. Acoustic fluidization (AF) is the possible physical process for the mechanical fault slip releasing seismic radiation through P-wave, S-wave, and surface waves. Several pre-seismic phenomena (SES, TEC, IR, EQL, and Radon gas) could be explained considering accumulation of electrical potential and associated electrical field in the hypocenter. This finding is a way forward to understand the true origin of an earthquake and paves the way for predicting an impending earthquake.

**Acknowledgments**

**Author contributions:** A. D. conceived the idea and wrote the manuscript. **Competing interests:** Authors declare no competing interest. **ORCID:** Atanu Das:0000-0002-1733-626X. **Data Statement:** All data are reported in the paper or in the supplementary materials. **Dedication**: Author (A.D) would like to dedicate this research article to his parents (Kananbala Das and Nagendranath Das), who have been a constant source of inspiration and have given him the drive and discipline to tackle any task with enthusiasm and determination.

**Supplementary Materials:**

Figures S1-S2

Tables S1-S5

External Database S1

Additional Supplementary Information

References (*1-53*)

**References and Notes:**

1. H. Kanamori, Magnitude scale and quantification of earthquakes. *Tectonophysics* **93**, 185-199 (1983).
2. T. C. Hanks, H. Kanamori, A moment magnitude scale. *Journal of Geophysical Research: Solid Earth* **84**, 2348-2350 (1979).
3. B. Gutenberg, C. F. Richter, Magnitude and energy of earthquakes. *Annals of Geophysics; Vol 53, No 1 (2010)*, (2010).
4. A. Skelton *et al.*, Hydrochemical Changes Before and After Earthquakes Based on Long-Term Measurements of Multiple Parameters at Two Sites in Northern Iceland—A Review. *Journal of Geophysical Research: Solid Earth* **124**, 2702-2720 (2019).
5. D. Ghosh, A. Deb, R. Sengupta, Anomalous radon emission as precursor of earthquake. *Journal of Applied Geophysics* **69**, 67-81 (2009).
6. S. P. Dimitar Ouzounov, Alexey Romanov, Alexander Romanov, Konstantin Tsybulya, Dimitri Davidenko, Menas Kafatos, Patrick Taylor, Atmosphere-Ionosphere Response to the M9 Tohoku Earthquake Revealed by Joined Satellite and Ground Observations. Preliminary results. *arXiv*, (2011).
7. A. K. Saraf *et al.*, Satellite detection of earthquake thermal infrared precursors in Iran. *Natural Hazards* **47**, 119-135 (2008).
8. A. K. Saraf, S. Choudhury, Cover: NOAA-AVHRR detects thermal anomaly associated with the 26 January 2001 Bhuj earthquake, Gujarat, India. *International Journal of Remote Sensing* **26**, 1065-1073 (2005).
9. M. Zhong *et al.*, Thermal Infrared and Ionospheric Anomalies of the 2017 Mw6.5 Jiuzhaigou Earthquake. *Remote Sensing* **12**, 2843 (2020).
10. S. UYEDA, in *A Critical Review of Van*. pp. 3-28.
11. A. Das, pH-Sensitive Ultra-thin Oxide Liquid Metal System. *arXiv* **2007.09843**, (2020).
12. M. Pourbaix, Atlas of electrochemical equilibria in aqueous solutions. (1974).
13. W. Fu *et al.*, Graphene Transistors Are Insensitive to pH Changes in Solution. *Nano Letters* **11**, 3597-3600 (2011).

14. R. R. K. Reddy, A. Chadha, E. Bhattacharya, Porous silicon based potentiometric triglyceride biosensor. *Biosensors and Bioelectronics* **16**, 313-317 (2001).
15. N. H. Al-Hardan *et al.*, High Sensitivity pH Sensor Based on Porous Silicon (PSi) Extended Gate Field-Effect Transistor. *Sensors* **16**, (2016).
16. N. Zehfroosh, M. Shahmohammadi, S. Mohajerzadeh, High-Sensitivity Ion-Selective Field-Effect Transistors Using Nanoporous Silicon. *IEEE Electron Device Letters* **31**, 1056-1058 (2010).
17. C. C. Liu, B. C. Bocchicchio, P. A. Overmyer, M. R. Neuman, A palladium-palladium oxide miniature pH electrode. *Science* **207**, 188-189 (1980).
18. H. Hwang *et al.*, A role for subducted super-hydrated kaolinite in Earth's deep water cycle. *Nature geoscience* **10**, 947-953 (2017).
19. I. S. McLintock, The Elovich Equation in Chemisorption Kinetics. *Nature* **216**, 1204-1205 (1967).
20. H. A. Taylor, N. Thon, Kinetics of Chemisorption1. *Journal of the American Chemical Society* **74**, 4169-4173 (1952).
21. F.-C. Wu, R.-L. Tseng, R.-S. Juang, Characteristics of Elovich equation used for the analysis of adsorption kinetics in dye-chitosan systems. *Chemical Engineering Journal* **150**, 366-373 (2009).
22. H. K. a. R. E. Grim, *Clay mineral*. (Encyclopedia Britannica, 2014).
23. D. C. S. Bain, B.F. L., Wilson, M. J., Ed, *Clay mineralogy: Spectroscopy and Chemical Determinative Methods*. (Champman and Hall New York, USA, 1994).
24. T. S. Mahadevan, S. H. Garofalini, Dissociative Chemisorption of Water onto Silica Surfaces and Formation of Hydronium Ions. *The Journal of Physical Chemistry C* **112**, 1507-1515 (2008).
25. L.-W. Kuo, S.-R. Song, E.-C. Yeh, H.-F. Chen, Clay mineral anomalies in the fault zone of the Chelungpu Fault, Taiwan, and their implications. *Geophysical Research Letters* **36**, (2009).
26. J. Kameda, M. Shimizu, Role of Clay Minerals on Tsunamigenic Faulting During Large Earthquakes. *Journal of the Clay Science Society of Japan (in Japanese)* **54**, 105-113 (2016).
27. B. M. Carpenter, C. Marone, D. M. Saffer, Weakness of the San Andreas Fault revealed by samples from the active fault zone. *Nature geoscience* **4**, 251-254 (2011).
28. H. K. Gupta, A review of recent studies of triggered earthquakes by artificial water reservoirs with special emphasis on earthquakes in Koyna, India. *Earth-Science Reviews* **58**, 279-310 (2002).
29. N. Deichmann *et al.*, paper presented at the American Geophysical Union, Fall Meeting 2007, December 01, 2007 2007.
30. S. Ide, S. Yabe, Y. Tanaka, Earthquake potential revealed by tidal influence on earthquake size–frequency statistics. *Nature geoscience* **9**, 834-837 (2016).
31. P. Steer *et al.*, Earthquake statistics changed by typhoon-driven erosion. *Scientific Reports* **10**, 10899 (2020).
32. G. Michas, K. Pavlou, F. Vallianatos, G. Drakatos, Correlation Between Seismicity and Water Level Fluctuations in the Polyphyto Dam, North Greece. *Pure and Applied Geophysics* **177**, 3851-3870 (2020).
33. S. Priyadarshi, S. Kumar, A. K. Singh, Changes in total electron content associated with earthquakes (M > 5) observed from GPS station, Varanasi, India. *Geomatics, Natural Hazards and Risk* **2**, 123-139 (2011).
34. S. A. Pulinets, Physical mechanism of the vertical electric field generation over active tectonic faults. *Advances in Space Research* **44**, 767-773 (2009).
35. E. H. Lay, X.-M. Shao, C. S. Carrano, Variation in total electron content above large thunderstorms. *Geophysical Research Letters* **40**, 1945-1949 (2013).
36. A. G. S. V.I.Gorny, A.A.Tronin, B.V.Shilin, TERRESTRIAL OUTGOING INFRARED RADIATION AS AN INDICATOR OF SEISMIC ACTIVITY. *arXiv*, (2020).
37. D. Ouzounov, S. P., Menas C. Kafatos, and Patrick Taylor, in *Pre-Earthquake Processes*. (2018), pp. 259-274.
38. J. R. Michalski, M. D. Kraft, T. G. Sharp, L. B. Williams, P. R. Christensen, Emission spectroscopy of clay minerals and evidence for poorly crystalline aluminosilicates on Mars from Thermal Emission Spectrometer data. *Journal of Geophysical Research: Planets* **111**, (2006).
39. F. T. Freund *et al.*, Stimulated infrared emission from rocks: assessing a stress indicator. *e-Earth* **2**, 7 (2007).
40. N. E. Whitehead, p. Ulusoy, Origin of Earthquake Light Associated with Earthquakes in Christchurch, New Zealand, 2010-2011 %J Earth Sciences Research Journal. **19**, 113-119 (2015).
41. I. Pidchenko, S. Salminen-Paatero, J. Rothe, J. Suksi, Study of uranium oxidation states in geological material. *Journal of Environmental Radioactivity* **124**, 141-146 (2013).
42. P. Tosi, P. Sbarra, V. De Rubeis, Earthquake sound perception. *Geophysical Research Letters* **39**, (2012).
43. V. Kuznetsov, Shock-wave model of the earthquake and Poincaré quantum theorem give an insight into the aftershock physics. *E3S Web Conf.* **62**, (2018).

44. J. Chen, C. Elmi, D. Goldsby, R. Gieré, Generation of shock lamellae and melting in rocks by lightning-induced shock waves and electrical heating. **44**, 8757-8768 (2017).
45. M. A. Pasek, M. Hurst, A Fossilized Energy Distribution of Lightning. *Scientific Reports* **6**, 30586 (2016).
46. K. Okazaki, G. Hirth, Dehydration of lawsonite could directly trigger earthquakes in subducting oceanic crust. *Nature* **530**, 81-84 (2016).
47. F. Giacco, L. Saggese, L. de Arcangelis, E. Lippiello, M. Pica Ciamarra, Dynamic Weakening by Acoustic Fluidization during Stick-Slip Motion. *Physical Review Letters* **115**, 128001 (2015).
48. K. Xia, S. Huang, C. Marone, Laboratory observation of acoustic fluidization in granular fault gouge and implications for dynamic weakening of earthquake faults. *GeoChemistryGeoPhysicsGeoSystems* **14**, 1012-1022 (2013).
49. H. J. Melosh, Acoustic Fluidization: Can sound waves explain why dry rock debris appears to flow like a fluid in some energetic geologic events? *American Scientist* **71**, 158-165 (1983).
50. Y.-G. Li, K. Aki, D. Adams, A. Hasemi, W. H. K. Lee, Seismic guided waves trapped in the fault zone of the Landers, California, earthquake of 1992. *Journal of Geophysical Research: Solid Earth* **99**, 11705-11722 (1994).
51. D. Sornette, Mechanochemistry: an hypothesis for shallow earthquakes. *ArXiv* **arXiv:cond-mat/9807400**, (30 Jul 1998).
52. T.-L. Lin, C. A. Langston, Infrasound from thunder: A natural seismic source. **34**, (2007).
53. M. E. Kappus, F. L. Vernon, Acoustic signature of thunder from seismic records. *Journal of Geophysical Research: Atmospheres* **96**, 10989-11006 (1991).

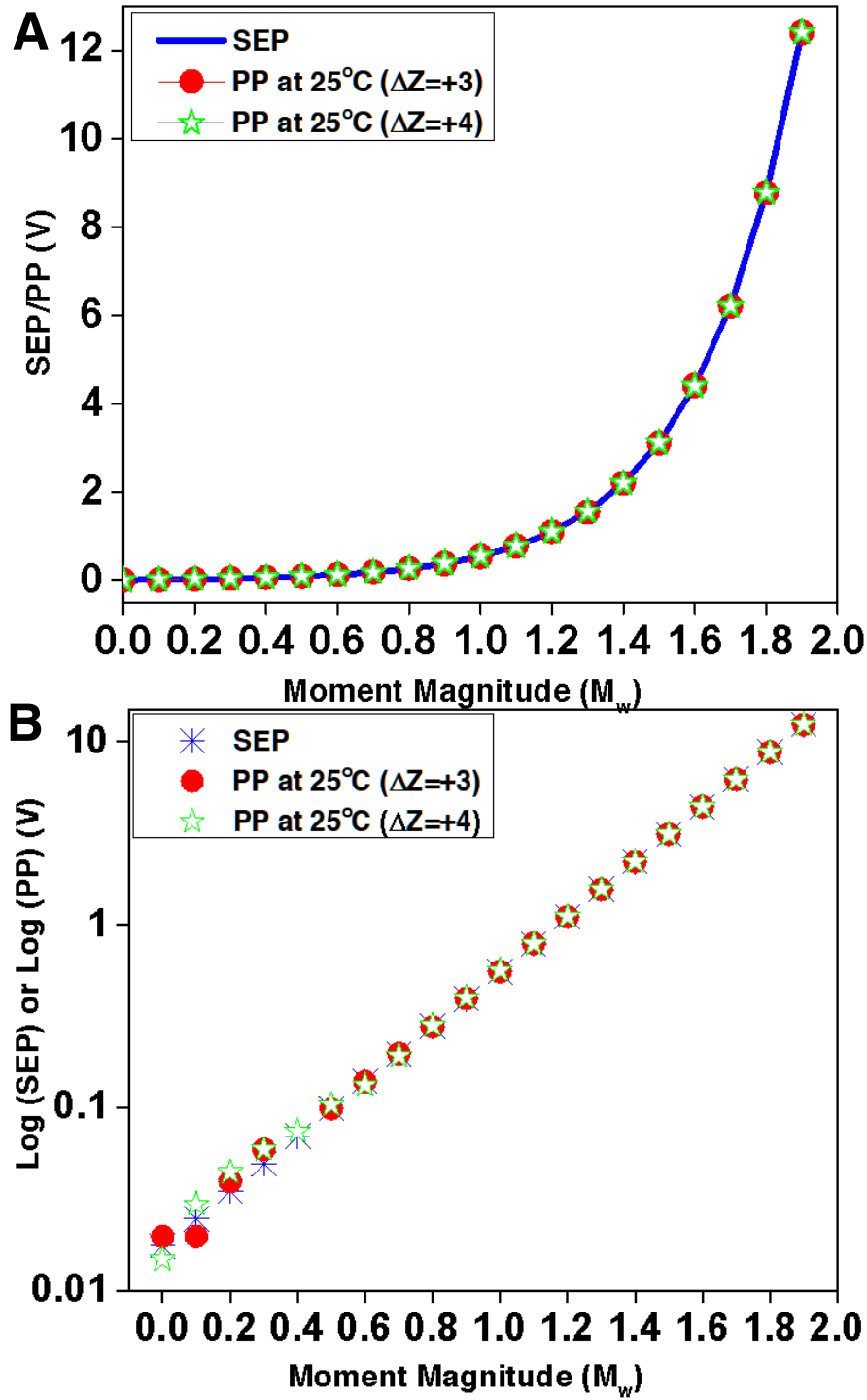

**Figure 1. Correlation between SEP and PP**
(A) Seismic electrical potential (SES) and Pourbaix Potential (PP) at 25°C vs. earthquake moment magnitude $M_w$ (0.0~1.9) plot. The PP in different ion exchange factors is closely matched with SEP values. (B) Semi-log plot of SEP(PP) vs. Earthquake moment magnitude $M_w$ (0.0~1.9) plot. The PP values are following the same trend as SEPs in an overlapped curve.

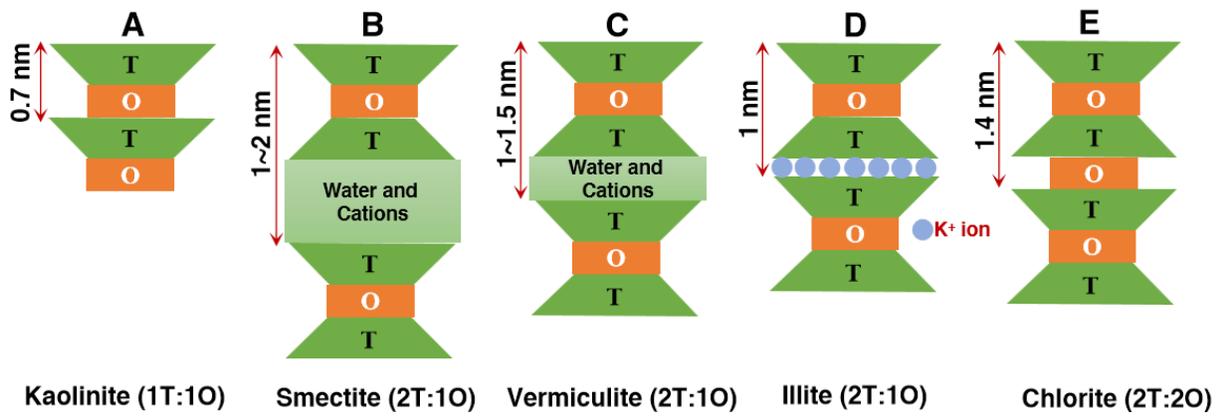

**Figure 2**. Different clay minerals are based on their layer stacking arrangements. Approximate basal layer spacing is indicated on the left of each type of clay minerals.

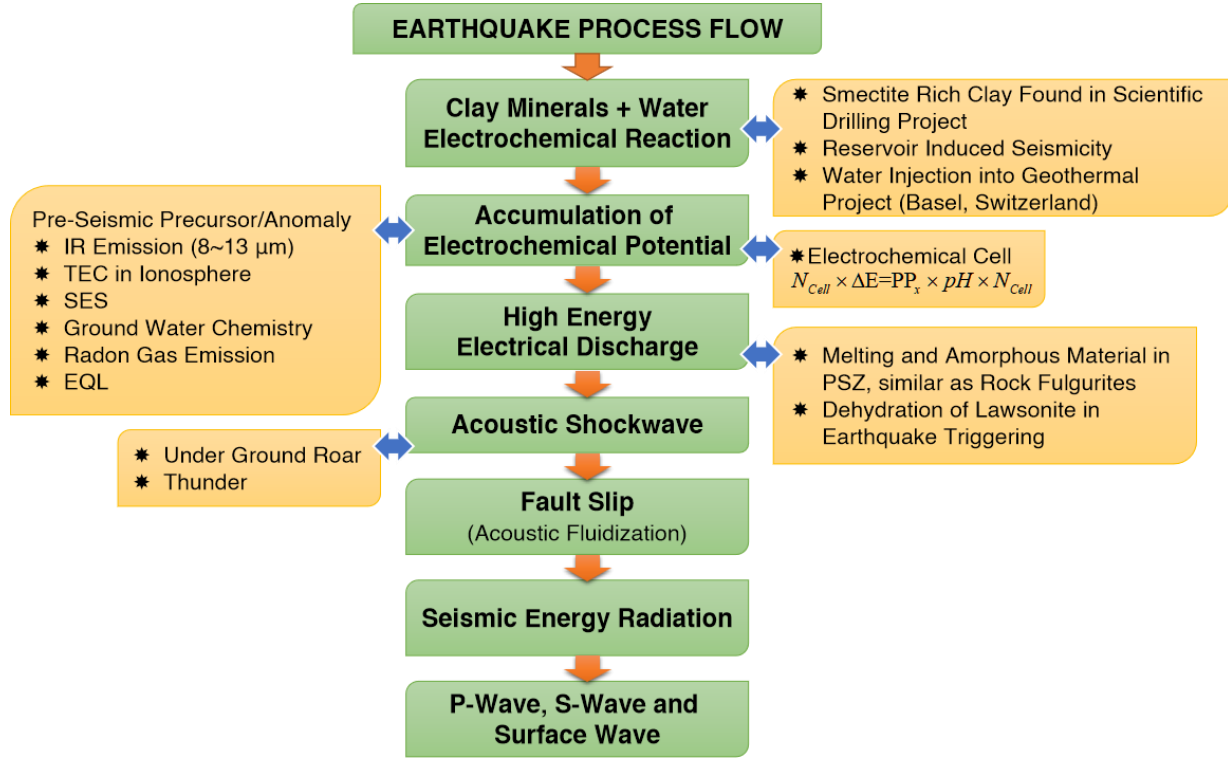

**Figure 3**. Schematic illustration of an earthquake process flow. The pre-seismic anomaly is linked to the energy accumulation process. Other evidence strengthens a few unique phenomena (i.e., electrochemical reaction, high energy electrical discharge, acoustic shock wave propagation) in an earthquake process.

**Table 1.** Details of Earthquake energy in Joule, Kilowatt-hour, and equivalent seismic electrical potential (SEP) (Joule/Coulomb)

| $M_w$ | Energy kJ ($\times 10^3$ J) | kWh ($\times 10^3$ Wh) | SEP $\times 3.6 \times 10^6$ V |
|---|---|---|---|
| 0 | 63.095 | 0.01752 | 0.01752 |
| 0.1 | 89.125 | 0.0247 | 0.0247 |
| 0.2 | 125.892 | 0.03497 | 0.03497 |
| 0.3 | 177.827 | 0.0493 | 0.0493 |
| 0.4 | 251.188 | 0.0697 | 0.0697 |
| 0.5 | 354.813 | 0.0985 | 0.0985 |
| 0.6 | 501.187 | 0.1392 | 0.1392 |
| 0.7 | 707.945 | 0.1966 | 0.1966 |
| 0.8 | 1000 | 0.2777 | 0.2777 |
| 0.9 | 1412.537 | 0.3923 | 0.3923 |
| 1 | 1995.262 | 0.5542 | 0.5542 |
| 1.1 | 2818.382 | 0.7828 | 0.7828 |
| 1.2 | 3981.071 | 1.1058 | 1.1058 |
| 1.3 | 5623.413 | 1.5620 | 1.5620 |
| 1.4 | 7943.282 | 2.2064 | 2.2064 |
| 1.5 | 11220.184 | 3.1167 | 3.1167 |
| 1.6 | 15848.931 | 4.4024 | 4.4024 |
| 1.7 | 22387.211 | 6.2186 | 6.2186 |
| 1.8 | 31622.776 | 8.7841 | 8.7841 |
| 1.9 | 44668.359 | 12.4078 | 12.4078 |

**Table 2.** Details of earthquake electrical potential (Joule/Coulomb), electrochemical potential from redox reaction of trivalent and tetravalent elemental oxide with different ion-exchange factors.

| $M_w$ | SEP | $\Delta Z=+3$ $x$ | $PP_x(V)$ | $\Delta Z=+4$ $x$ | $PP_x(V)$ |
|---|---|---|---|---|---|
| 0 | 0.01752 | 1/3 | 0.01972 | 1/4 | 0.01479 |
| 0.1 | 0.0247 | 1/3 | 0.01972 | 2/4 | 0.02958 |
| 0.2 | 0.03497 | 2/3 | 0.03944 | 3/4 | 0.04437 |
| 0.3 | 0.0493 | 3/3 | 0.05916 | 4/4 | 0.05916 |
| 0.4 | 0.0697 | 4/3 | 0.07888 | 5/4 | 0.07395 |
| 0.5 | 0.0985 | 5/3 | 0.09860 | 7/4 | 0.10353 |
| 0.6 | 0.1392 | 7/3 | 0.13804 | 9/4 | 0.133 |
| 0.7 | 0.1966 | 10/3 | 0.19720 | 13/4 | 0.19227 |
| 0.8 | 0.2777 | 14/3 | 0.27608 | 19/4 | 0.28101 |
| 0.9 | 0.3923 | 20/3 | 0.39440 | 27/4 | 0.39933 |
| 1 | 0.5542 | 28/3 | 0.55216 | 38/4 | 0.56202 |
| 1.1 | 0.7828 | 40/3 | 0.78880 | 53/4 | 0.78387 |
| 1.2 | 1.1058 | 56/3 | 1.10432 | 75/4 | 1.10925 |
| 1.3 | 1.5620 | 79/3 | 1.55788 | 106/4 | 1.56774 |
| 1.4 | 2.2064 | 112/3 | 2.20864 | 149/4 | 2.20371 |
| 1.5 | 3.1167 | 158/3 | 3.11576 | 211/4 | 3.12069 |
| 1.6 | 4.4024 | 224/3 | 4.41728 | 298/4 | 4.40742 |
| 1.7 | 6.2186 | 315/3 | 6.2118 | 420/4 | 6.2118 |
| 1.8 | 8.7841 | 445/3 | 8.77540 | 594/4 | 8.78526 |
| 1.9 | 12.4078 | 629/3 | 12.40388 | 839/4 | 12.40881 |

**Table 3.** Details of moment magnitude and corresponding SEP or (PP) and pH×$N_{Cell}$

| Magnitude ($M_w$) | SEP (V) or PP(V) | pH×$N_{Cell}$ | pH | $N_{Cell}$ |
|---|---|---|---|---|
| 0.0~1.9 | 0.01752~12.40 | $3.6 \times 10^6$ | 1~12 | $(3.6$~$0.3) \times 10^6$ |
| 2.0~3.9 | 0.01752~12.40 | $3.6 \times 10^9$ | 1~12 | $(3.6$~$0.3) \times 10^9$ |
| 4.0~5.9 | 0.01752~12.40 | $3.6 \times 10^{12}$ | 1~12 | $(3.6$~$0.3) \times 10^{12}$ |
| 6.0~7.9 | 0.01752~12.40 | $3.6 \times 10^{15}$ | 1~12 | $(3.6$~$0.3) \times 10^{15}$ |
| 8.0~9.9 | 0.01752~12.40 | $3.6 \times 10^{18}$ | 1~12 | $(3.6$~$0.3) \times 10^{18}$ |

**Table 4.** Comparison of a few important properties of clay minerals. The CEC and specific surface area of smectite and vermiculite are higher than all other clay minerals.

| Properties | Kaolinite | Smectite | Vermiculite | Illite | Chlorite |
|---|---|---|---|---|---|
| Expansion | No | High | Moderate | No | No |
| Substitution | Tetrahedral | Octahedral | Tetrahedral | Tetrahedral | Octahedral |
| CEC (meq/100 gm) | 3-15 | 60-150 | 100-150 | 10-40 | 20-40 |
| Surface Area $m^2/gm$ | 5-20 | 700-800 | 300-500 | 80-100 | 80 |
| Reactive Surface and Adsorption | External Surface and Edges | External Surface and Interlayer spacing | External Surface and Interlayer spacing | External Surface and Edges | External Surface and Edges |

# Supplementary Materials for

## Earthquake and Electrochemistry: Unraveling the Unpredictable

**Author:** Atanu Das[1, *]

*Correspondence to: atanu@gmx.com (Atanu Das*); atanu.esensex@protonmail.com

**Supplementary Materials**
    Supplementary Text
    Figs. S1 to S2
    Tables S1-S5
    Separate Excel Datasheet
**Other Supplementary Materials for this manuscript include the following:**
    Additional Information: Evidence from Eyewitness

**References:** (1-53)

**Demonstration of Electrochemical Cell using Liquid metal Potentiometric Sensor**

Recently A. Das et al. (*11*) demonstrated a potentiometric sensor using an ultrathin $Ga_2O_3$-liquid metal system. The potentiometric sensor is fundamentally inherited from the electrochemical cell, where one half cell reaction occurs in a working electrode-aqueous interface. Another half-cell reaction is a well understood non-interfering reaction occurring at the reference (Ag|AgCl) electrode. Here we demonstrated that the half-cell reaction in each working electrode (LMPD) could be added like a regular electrochemical cell (Figure S1). The measured potential for one LMPD is -1.100 V in pH 10 aqueous buffer solution where the voltage is doubled (-2.200 V) for properly connected two LMPD electrodes in series. For **N** number cells ($N_{Cell}$), simply the resultant potential will be $N_{Cell} \times 1.100$ V. The measured potential from a single cell (-1.100 V) can be understood from the following electrochemical reaction and associated electrode potential. The possible electrochemical response at the LMPD/aqueous pH buffer (pH 10) solution as follows,

$$Ga + 3H_2O \leftrightarrow GaO_3^{3-} + 6H^+ + 3e^- \qquad (1)$$

The electrode potential for the above reaction will be,

$$E = E^0 - 0.05916\left(\frac{6}{3}\right)pH - 0.0197\log(GaO_3^{3-}) \qquad (2)$$

The unified Nernst equation for electrode potential including Pourbaix factor(*11*) will be

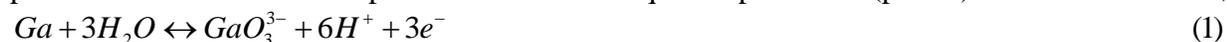

$$E = E^0 - 0.0591 \times pb \times \frac{6}{3} \times \log(a_{H^+}) \quad [x = \frac{6}{3}] \qquad (3)$$

$$\Delta E = E - E^0 = -0.0591 \times 0.932 \times 2 \times 10 = -1.100 \quad [pb=0.932] \qquad (4)$$

We could produce a sizeable voltage from the electrochemical potential of several similar LMPD electrode/aqueous pH buffer systems.

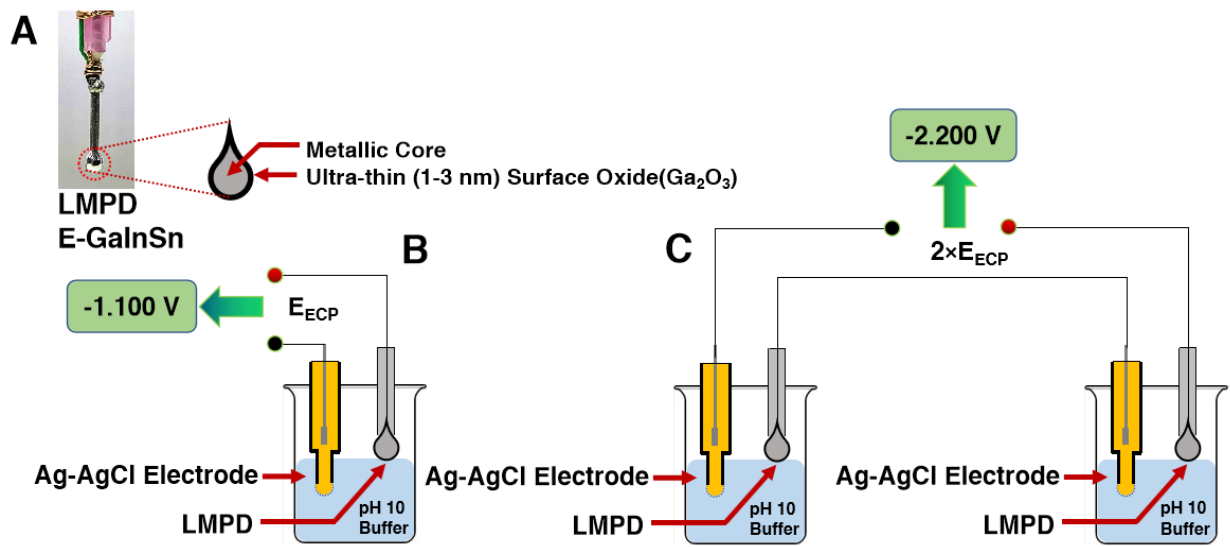

**Figure S1.**
Demonstration of the electrochemical cell using liquid metal pendant drop (LMPD). (A) Photographs of LMPD (Eutectic GaInSn, 99.99% pure) consist of the inner metallic core and ultrathin surface oxide. (B) The electrochemical potential is generated in the interfacial reaction with pH 10 aqueous buffer and measured electrochemical cell potential ($E_{ECP}$) is -1.100 V for single LMPD configuration, (C) Electrochemical half-cell reaction in LMPD probe could be added like regular electrochemical cell and measured $E_{ECP}$ for two cells in series is -2.200 V.

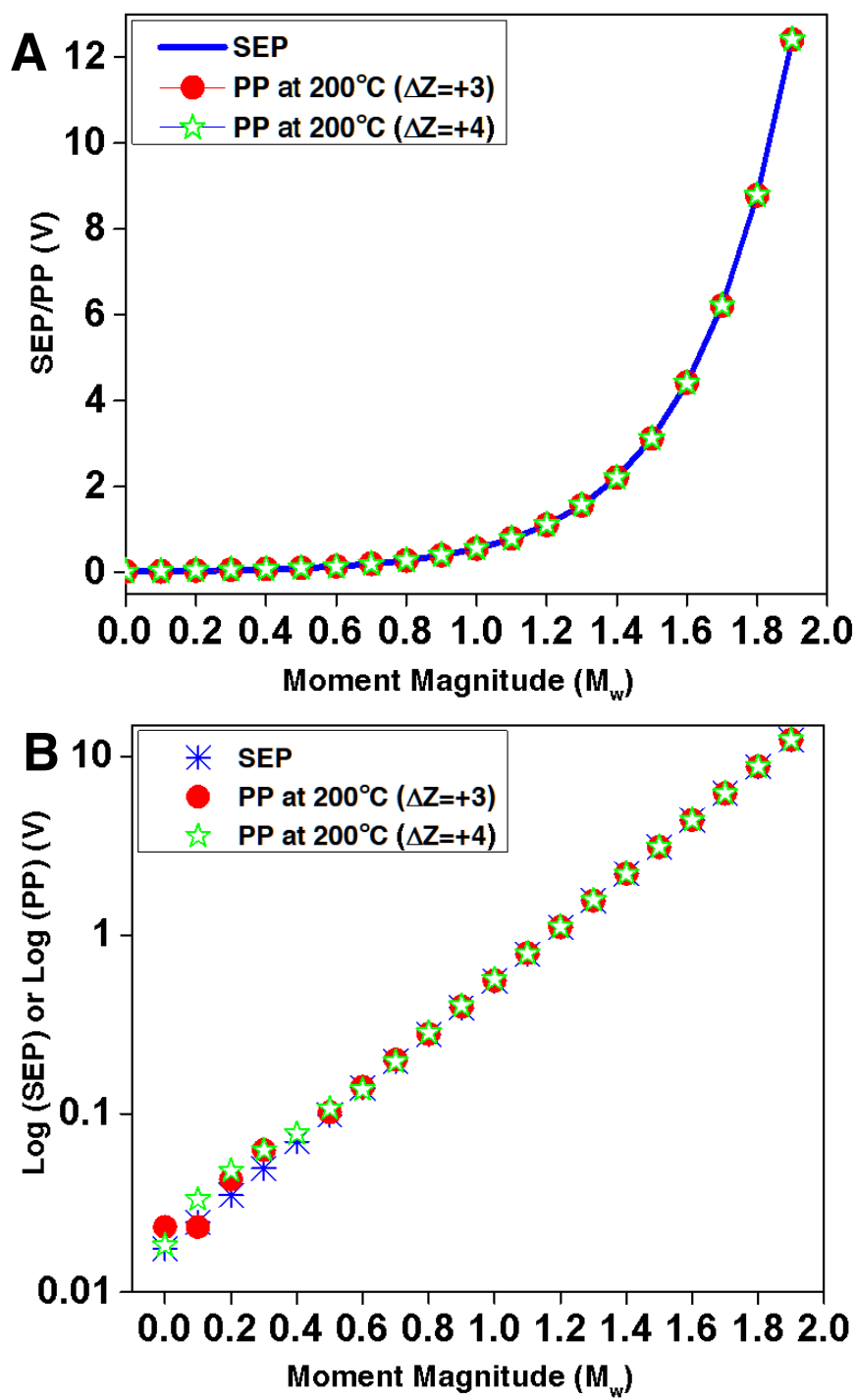

**Figure S2.**
(A)Seismic electrical potential and Pourbaix Potential at 200°C vs. Earthquake moment magnitude $M_w$ (0.0~1.9) plot. PP in different ion exchange factors are closely match with SEP values. (B)Semi-log plot of SEP(PP) vs. Earthquake moment magnitude $M_w$ (0.0~1.9).

**Table S1.** Details of Earthquake energy in Joule, Kilowatt-hour, and equivalent seismic electrical potential (Joule/Coulomb) in $M_w$ (2.0~3.9).

| $M_w$ | Energy MJ($\times 10^6$J) | MWh ($\times 10^6$W h) | SEP ($\times 3.6 \times 10^9$ V) |
|---|---|---|---|
| 2.0 | 63.095 | 0.01752 | 0.01752 |
| 2.1 | 89.125 | 0.0247 | 0.0247 |
| 2.2 | 125.892 | 0.03497 | 0.03497 |
| 2.3 | 177.827 | 0.0493 | 0.0493 |
| 2.4 | 251.188 | 0.0697 | 0.0697 |
| 2.5 | 354.813 | 0.0985 | 0.0985 |
| 2.6 | 501.187 | 0.1392 | 0.1392 |
| 2.7 | 707.945 | 0.1966 | 0.1966 |
| 2.8 | 1000 | 0.2777 | 0.2777 |
| 2.9 | 1412.537 | 0.3923 | 0.3923 |
| 3 | 1995.262 | 0.5542 | 0.5542 |
| 3.1 | 2818.382 | 0.7828 | 0.7828 |
| 3.2 | 3981.071 | 1.1058 | 1.1058 |
| 3.3 | 5623.413 | 1.5620 | 1.5620 |
| 3.4 | 7943.282 | 2.2064 | 2.2064 |
| 3.5 | 11220.184 | 3.1167 | 3.1167 |
| 3.6 | 15848.931 | 4.4024 | 4.4024 |
| 3.7 | 22387.211 | 6.2186 | 6.2186 |
| 3.8 | 31622.776 | 8.7841 | 8.7841 |
| 3.9 | 44668.359 | 12.4078 | 12.4078 |

**Table S2.** Details of Earthquake energy in Joule, Kilowatt-hour, and equivalent seismic electrical potential (Joule/Coulomb) in $M_w$ (4.0~5.9).

| $M_w$ | Energy GJ($\times 10^9$J) | GWh ($\times 10^9$W h) | SEP ($\times 3.6 \times 10^{12}$ V) |
|---|---|---|---|
| 4.0 | 63.095 | 0.01752 | 0.01752 |
| 4.1 | 89.125 | 0.0247 | 0.0247 |
| 4.2 | 125.892 | 0.03497 | 0.03497 |
| 4.3 | 177.827 | 0.0493 | 0.0493 |
| 4.4 | 251.188 | 0.0697 | 0.0697 |
| 4.5 | 354.813 | 0.0985 | 0.0985 |
| 4.6 | 501.187 | 0.1392 | 0.1392 |
| 4.7 | 707.945 | 0.1966 | 0.1966 |
| 4.8 | 1000 | 0.2777 | 0.2777 |
| 4.9 | 1412.537 | 0.3923 | 0.3923 |
| 5 | 1995.262 | 0.5542 | 0.5542 |
| 5.1 | 2818.382 | 0.7828 | 0.7828 |
| 5.2 | 3981.071 | 1.1058 | 1.1058 |
| 5.3 | 5623.413 | 1.5620 | 1.5620 |
| 5.4 | 7943.282 | 2.2064 | 2.2064 |
| 5.5 | 11220.184 | 3.1167 | 3.1167 |
| 5.6 | 15848.931 | 4.4024 | 4.4024 |
| 5.7 | 22387.211 | 6.2186 | 6.2186 |
| 5.8 | 31622.776 | 8.7841 | 8.7841 |
| 5.9 | 44668.359 | 12.4078 | 12.4078 |

**Table S3.** Details of Earthquake energy in Joule, Kilowatt-hour, and equivalent seismic electrical potential (Joule/Coulomb) in $M_w$ (6.0~7.9).

| $M_w$ | Energy TJ($\times 10^{12}$J) | TWh ($\times 10^{12}$W h) | SEP ($\times 3.6 \times 10^{15}$ V) |
|---|---|---|---|
| 6.0 | 63.095 | 0.01752 | 0.01752 |
| 6.1 | 89.125 | 0.0247 | 0.0247 |
| 6.2 | 125.892 | 0.03497 | 0.03497 |
| 6.3 | 177.827 | 0.0493 | 0.0493 |
| 6.4 | 251.188 | 0.0697 | 0.0697 |
| 6.5 | 354.813 | 0.0985 | 0.0985 |
| 6.6 | 501.187 | 0.1392 | 0.1392 |
| 6.7 | 707.945 | 0.1966 | 0.1966 |
| 6.8 | 1000 | 0.2777 | 0.2777 |
| 6.9 | 1412.537 | 0.3923 | 0.3923 |
| 7 | 1995.262 | 0.5542 | 0.5542 |
| 7.1 | 2818.382 | 0.7828 | 0.7828 |
| 7.2 | 3981.071 | 1.1058 | 1.1058 |
| 7.3 | 5623.413 | 1.5620 | 1.5620 |
| 7.4 | 7943.282 | 2.2064 | 2.2064 |
| 7.5 | 11220.184 | 3.1167 | 3.1167 |
| 7.6 | 15848.931 | 4.4024 | 4.4024 |
| 7.7 | 22387.211 | 6.2186 | 6.2186 |
| 7.8 | 31622.776 | 8.7841 | 8.7841 |
| 7.9 | 44668.359 | 12.4078 | 12.4078 |

**Table S4.** Details of Earthquake energy in Joule, Kilowatt-hour, and equivalent seismic electrical potential (Joule/Coulomb) in $M_w$ (8.0~9.9).

| $M_w$ | Energy PJ($\times 10^{15}$J) | PWh ($\times 10^{15}$W h) | SEP ($\times 3.6 \times 10^{18}$ V) |
|---|---|---|---|
| 8.0 | 63.095 | 0.01752 | 0.01752 |
| 8.1 | 89.125 | 0.0247 | 0.0247 |
| 8.2 | 125.892 | 0.03497 | 0.03497 |
| 8.3 | 177.827 | 0.0493 | 0.0493 |
| 8.4 | 251.188 | 0.0697 | 0.0697 |
| 8.5 | 354.813 | 0.0985 | 0.0985 |
| 8.6 | 501.187 | 0.1392 | 0.1392 |
| 8.7 | 707.945 | 0.1966 | 0.1966 |
| 8.8 | 1000 | 0.2777 | 0.2777 |
| 8.9 | 1412.537 | 0.3923 | 0.3923 |
| 9 | 1995.262 | 0.5542 | 0.5542 |
| 9.1 | 2818.382 | 0.7828 | 0.7828 |
| 9.2 | 3981.071 | 1.1058 | 1.1058 |
| 9.3 | 5623.413 | 1.5620 | 1.5620 |
| 9.4 | 7943.282 | 2.2064 | 2.2064 |
| 9.5 | 11220.184 | 3.1167 | 3.1167 |
| 9.6 | 15848.931 | 4.4024 | 4.4024 |
| 9.7 | 22387.211 | 6.2186 | 6.2186 |
| 9.8 | 31622.776 | 8.7841 | 8.7841 |
| 9.9 | 44668.359 | 12.4078 | 12.4078 |

**Table S5.** Details of Seismic Electrical Potential (SEP) in $M_w$ (0~1.9), Pourbaix Potential (PP) for different ionic exchange factor at 200ºC for material with $\Delta Z=+3$ and $+4$

| $M_w$ | SEP (V) | $\Delta Z=+3$ $x$ | $PP_x$ (V) (200ºC) | $\Delta Z=+4$ $x$ | $PP_x$ (V) (200ºC) |
|---|---|---|---|---|---|
| 0 | 0.01752 | 1/3 | 0.02322 | 1/4 | 0.01829 |
| 0.1 | 0.0247 | 1/3 | 0.02322 | 2/4 | 0.03308 |
| 0.2 | 0.03497 | 2/3 | 0.04294 | 3/4 | 0.04787 |
| 0.3 | 0.0493 | 3/3 | 0.06266 | 4/4 | 0.06266 |
| 0.4 | 0.0697 | 4/3 | 0.08238 | 5/4 | 0.07745 |
| 0.5 | 0.0985 | 5/3 | 0.1021 | 7/4 | 0.10703 |
| 0.6 | 0.1392 | 7/3 | 0.14154 | 9/4 | 0.1365 |
| 0.7 | 0.1966 | 10/3 | 0.2007 | 13/4 | 0.19577 |
| 0.8 | 0.2777 | 14/3 | 0.27958 | 19/4 | 0.28451 |
| 0.9 | 0.3923 | 20/3 | 0.3979 | 27/4 | 0.40283 |
| 1 | 0.5542 | 28/3 | 0.55566 | 38/4 | 0.56552 |
| 1.1 | 0.7828 | 40/3 | 0.7923 | 53/4 | 0.78737 |
| 1.2 | 1.1058 | 56/3 | 1.10782 | 75/4 | 1.11275 |
| 1.3 | 1.5620 | 79/3 | 1.56138 | 106/4 | 1.57124 |
| 1.4 | 2.2064 | 112/3 | 2.21214 | 149/4 | 2.20721 |
| 1.5 | 3.1167 | 158/3 | 3.11926 | 211/4 | 3.12419 |
| 1.6 | 4.4024 | 224/3 | 4.42078 | 298/4 | 4.41092 |
| 1.7 | 6.2186 | 315/3 | 6.2153 | 420/4 | 6.2153 |
| 1.8 | 8.7841 | 445/3 | 8.7789 | 594/4 | 8.78876 |
| 1.9 | 12.4078 | 629/3 | 12.40738 | 839/4 | 12.41231 |

## Additional Information

Earthquake sounds like thunder. Several incidents are scripted below.

**Evidence-1**
[https://www.scientificamerican.com/article/drilling-for-earthquakes/]
Editor's Note (11/7/2016): A magnitude 5.0 earthquake struck central Oklahoma on Sunday, Nov. 6, damaging buildings across the area. Experts have not yet tied the quake to oil and gas production, but the epicenter is near wastewater disposal wells, structures that have been implicated in earlier quakes in the state.
To Cathy Wallace, the earthquakes that have been rattling her tidy suburban home in Dallas feel like ***underground thunderstorms***. First comes a distant roar, then a boom and a jolt. Her house shakes, and the windows shudder. Framed prints on the walls clatter and tilt. A heavy glass vase tips over with a crash.
The worst moments are the ones between the rumble and the impact. "Every time it happens you know it's going to hit, but you don't know how severe it's going to be," she says. "Is this going to be a bigger one? Is this the part where my house falls down? It's scary. It's very scary."

**Evidence-2**
[https://www.youtube.com/watch?v=q1wg2IbA0oo]
In this recording of the 2011 Japanese earthquake, taken from measurements in California, the quake created subtle movements deep in the San Andreas Fault. The initial noise, which sounds like ***distant thunder***, corresponds with the Japanese mainshock. Afterwards, a continuous high-pitch sound, similar to rainfall that turns on and off, represents induced tremor activity at the fault. This animation not only help scientists explain the concept of distant triggering to general audiences, but also provides a useful tool for researchers to better identify and understand such seismic signals in other regions. Georgia Tech Associate Professor Zhigang Peng has converted the seismic waves from last year's earthquakes into audio files. The results allow experts and general audiences to "hear" what the quake sounded like as it moved through the earth and around the globe.

**Evidence-3**
[https://news.berkeley.edu/2018/01/12/transcript-berkeley-seismologists-capture-earthquakes-rumble/]
There are a bunch of sensors all over campus, but the sensors that captured this quake are in a seismic station in the Byerly Vault about 140 feet underground in the Berkeley hills behind the UC Botanical Garden.
This earthquake's epicenter was near the Claremont Hotel in Berkeley, about 8 miles below the earth's surface. The sound it produced took about two seconds to get to the Byerly Vault.
So, what are we hearing?
Hellweg: "Hah, an earthquake."
Peggy Hellweg and her colleagues at Berkeley's Seismological Lab monitor the sensors.
Hellweg: "To a certain extent, ***earthquakes are like thunder***. Something happens and sound waves go out."

**Evidence-4**
[https://www.huffingtonpost.com.au/2017/09/01/extremely-loud-thunder-shakes-adelaide-with-earthquake-like-force_a_23194392/]
*'Extremely Loud Thunder'* Shakes Adelaide With Earthquake-Like Force

**Evidence-5**
[https://www.volcanodiscovery.com/earthquakes/quake-info/5903826/mag3quake-Sep-9-2020-2-Km-SSE-of-East-Freehold-New-Jersey-USA.html]
Minor mag. 3.1 earthquake - 2 Km SSE of East Freehold, New Jersey (USA) on Wednesday, 9 September 2020
Many people reported *__thunder like rumble__*, explosion as comment section in the article.

**Evidence-6**
https://magicvalley.com/news/local/skyquake-science-theres-little-research-into-the-mystery/article_9b63067b-34a7-5713-888e-8675d4bdb6ad.html
Shallow quake, *__sonic boom and rumble__*
Skyquake?

**Evidence-7**
https://www.tandfonline.com/doi/pdf/10.1080/00288306.1968.10423671
1844 Nov22 1h± Riwaka D
"two slight shocks ... which lasted about a minute ... accompanied by a *__distinct rumbling noise resembling thunder__*".

**Evidence-8**
https://researchportal.helsinki.fi/fi/publications/shallow-swarm-type-earthquakes-in-south-eastern-finland
"The swarm has been widely felt and reports on *__explosion- or thunder-like sounds__* accompanied by shaking of the ground are common"

**Evidence-9**
[https://www.state.nj.us/dep/njgs/enviroed/freedwn/njequakes1977.pdf]
Eyewitness of four incidents were scripted in the article (link of the article provided)
1727, November 9
"On the twenty-ninth day of October," (November 9 Gregorian Calendar,) "between ten and eleven, It was a terrible earthquake. It came with a dreadful roaring, as If *__It was thunder__*, and then a pounce like grate guns two or three times close one after another.
"1755, November 18   The earthquake came with a roaring sound like *__distant thunder__*, seemingly from the northwest. The shock resembled a long roiling sea -15- and It was necessary to hold something to prevent being thrown to the ground."
1870. October 20 Centered near Bale-St.Paul, Quebec
"A hasty note to let you know the disasters that were suddenly caused, here and In the vicinity, by the strangest earthquake In the memory of man. Approximately half an hour before noon a *__thunderbolt__* - this Is the only word I can use for what happened - an enormous detonation threw everyone in a state of shock and the earth started not to shake but to boll In a manner to cause dizziness not only to the people In houses but also to the ones In the open air.

1872, July 11 The shock was felt over a nearly circular area, 10 mi les In diameter with New Rochelle, N.Y., at the center. At Port Washington, L.I., long pendulum clocks stopped. ***A rumbling noise, then sounds like bursts of thunder, accompanied the shock***. Felt in New Jersey.

**Evidence-10**
[https://www.csmonitor.com/Science/2015/0228/Mysterious-booms-shake-Washington-state-community]
Mysterious booms shake Washington state community
"One resident, Michelle Kaake, heard ***two booms at her home***. The first vibrated the floor and rattled the windows. She heard another one about five minutes later."
***Shallow earthquakes can cause loud booms***, according to the U.S. Geological Survey website. "No one knows for sure, but scientists speculate that these 'booms' are probably small shallow earthquakes that are too small to be recorded," the USGS website said.
The USGS later reported that the mysterious booms that shook the town were the result of a 1.5-magnitude earthquake. Although not a huge event, the earthquake caused a swarm of several small quakes in a short time. Paul Caruso, a geophysicist from the USGS, told the local Fox News affiliate that most people wouldn't normally feel a 1.5-magnitude earthquake, but that the rock in Wisconsin is very old and well consolidated, allowing residents to feel otherwise sensitive rumblings.

**Evidence-11**
[https://apps.peer.berkeley.edu/education/files/USAToday_Earthquake.pdf]
"It was like a ***freight train*** coming through," says Larry Lafaive, an emergency services dispatcher in Plattsburgh who had reported to work right before the quake struck. "It kept getting louder and louder and louder, and the whole building started shaking. We didn't know what it was, and then someone said, 'Hey, we're having an earthquake.' " With a ***rumbling sound like a freight train***, a weekend earthquake reminded the Northeast that seismic events aren't just for California. The quake, which struck Upstate New York at 6:50 a.m. Saturday, sent tremors from Maine to Maryland. It had a magnitude of 5.1, according to the National Earthquake Information Center of the U.S. Geological Survey. NEWS - MONDAY - April 22, 2002 – 7A By Martha T. Moore USA TODAY